\documentclass[10pt,conference]{IEEEtran}
\IEEEoverridecommandlockouts
\usepackage{graphicx} % Required for inserting images

\usepackage{xspace}
\usepackage{algorithm}  
\usepackage{algorithmic}
\usepackage{enumitem}

\usepackage{url} 
\usepackage{mathbbol}
\newcommand{\sol}{{\em NetScaNDN}\xspace}
\usepackage{xcolor}
\usepackage{amsmath}
\usepackage{caption}

\usepackage[caption=false,font=normalsize,labelfont=sf,textfont=sf]{subfig}

\newcounter{magicrownumbers}

\begin{document}

\author{\IEEEauthorblockN{Amir Esmaeili}
\IEEEauthorblockA{\textit{Department of Computer Science} \\
\textit{Binghamton University}\\
Binghamton, NY, USA \\
aesmaeili@binghamton.edu}
%\and
%\IEEEauthorblockN{Abderrahmen Mtibaa}
%\IEEEauthorblockA{\textit{Department of Computer Science} \\
%\textit{University of Missouri St. Louis}\\
%St. Louis, MO, USA \\
%amtibaa@umsl.edu}
\and
\IEEEauthorblockN{Maryam Fazli}
\IEEEauthorblockA{\textit{Department of Computer Science} \\
\textit{Binghamton University}\\
Binghamton, NY, USA\\
mfazli@binghamton.edu}
}

\title{ \sol: A Scalable  and Flexible Testbed To Evaluate NDN on Multiple Infrastructures}
\maketitle

\begin{abstract}
 The evolution from traditional IP-based networking to Named Data Networking (NDN) represents a paradigm shift to address the inherent limitations of current network architectures, such as scalability, mobility, and efficient data distribution. NDN introduces an information-centric approach where data is identified and retrieved based on names rather than locations, offering more efficient data dissemination and enhanced security. However, the transition to NDN, alongside the need to integrate it with existing IP infrastructures, necessitates the development of flexible and scalable testbeds that support diverse experimental scenarios across various physical media and networking protocol stacks.

In this paper, we present \sol, a scalable, flexible, and plug-and-play testbed designed to facilitate such experiments. \sol employs an automated process for node discovery, configuration, and installation, enabling seamless setup and execution of experiments on both wired and wireless infrastructures simultaneously. Additionally, it incorporates a central log repository using the syslog protocol, allowing comprehensive measurement and evaluation of user-defined metrics across different network layers. \sol offers a robust platform for researchers to explore and validate various networking scenarios, advancing the study of IP and NDN-based applications.
\end{abstract}


\section{Introduction}

In the realm of modern networking, the evolution from traditional IP-based networking to Name Data Networking (NDN)~\cite{zhang2014named} represents a paradigm shift aimed at addressing the limitations of current network architectures like mobility, load balancing, and content centric networking. IP networking, which has been the backbone of the internet for decades, relies on location-based addressing and host to host communications where data retrieval is inherently tied to specific network addresses. This model, while effective, faces challenges in scalability, mobility, and efficient data distribution, especially as the number of connected devices continues to grow exponentially. 

Named Data Networking, on the other hand, proposes an information-centric approach where data is identified and retrieved based on names rather than locations. This fundamental shift enables more efficient data dissemination, inherent support for mobility, and enhanced security mechanisms. However, the transition to NDN and the need to integrate it with existing IP infrastructures pose significant challenges, necessitating flexible, and scalable testbeds that can support diverse experimental scenarios in multiple physical medium, and networking protocol stacks.

Although some researchers implement a general testbed for specific physical medium~\cite{bardhi2022sim2testbed,ni2015named,nurkahfi2021design}, network protocol stack~\cite{lim2018ndn}, or network-based applications~\cite{lim2018ndn}, nodes scalability, flexibility of implementation, and automatic node configuration which is important in large scale scenarios still is not well studied. There are some works that designed a testbed on specific network infrastructure like ISM band wireless~\cite{saxena2017implementation}, Cellular 4G infrastructure~\cite{4G2015testbed,LTE2019testbed}, edge network~\cite{li2022icast}, a Virtual Machine(VM) environment~\cite{bardhi2022sim2testbed}, moreover Lim testbed~\cite{lim2018ndn} is an infrastructure for climate changes data dissemination which is a new data specific testbed, however, a flexible testbed should be able to run different experiments on various mediums, and different networking protocol stacks, with capability of running multiple applications.

On the other hand, scalability is an important factor for a testbed, while testbed users should be able to scale several nodes, network topology, and their configurations. Some testbeds exploit the scale-out feature of virtual environments~\cite{bardhi2022sim2testbed}, however, virtual networks are a kind of application layer which artificially simulate real networks, furthermore, they can not implement real-world wireless infrastructure. 

In this paper, we present \sol a scalable, flexible, and plug-and-play testbed that exploits an automated process to discover, configure, and install all requirements on the nodes, in addition, \sol can run experiments on two different wired, and wireless infrastructures simultaneously. \sol exploits a central log repository for all nodes that store raw logs from the physical layer, up to the application layer, so it can measure several user-defined metrics for evaluation of running experiments. Compared to state-of-the-art solutions, \sol is a scalable testbed that can run a variety of experiments in multiple mediums and network protocol stack (IP, and NDN) on complicated network topologies, also the automated process of configuration makes \sol a very flexible testbed. The evaluation module of our test bed  shows that it can run complicated scenarios in a reasonable time. 

 All contributions of this paper are summarized as following lines: 
\begin{itemize}
    \item \sol is a general multipurpose testbed for IP and NDN applications run different experiments in wired, and wireless mediums. Unlike state of the art, it can run experiments concurrently on all mediums, and network protocol stacks (IP and NDN). 
    \item We also propose a fully automated process to discover nodes, install required software, and packages, and configure the network, and applications on each node to run for experiments. This process can be run in all Operating Systems (OS) regardless of hardware dependency, and it makes our testbed scalable and flexible to run experiments with several nodes, different software, and various network protocols.
    \item To evaluate different metrics of running experiments on the testbed with independent nodes, we use a central log repository based on syslog protocol to store raw logs, any required metrics can be defined and measured in this repository. Moreover, we run multiple complicated networking scenarios in our testbed.   
\end{itemize}

The rest of this paper is categorized as follows: In section~\ref{related works} we explain about state of art and related works, then in section~\ref{NDN} we explain the NDN technology and its main features, later in section~\ref{testbed overview} we discuss the testbed overview, included modules and their implementation. Finally, the conclusion and future direction will be discussed in section~\ref{conclusion}.

\section{Related works}\label{related works}

In the NDN workspace, several papers implement either a general~\cite{bardhi2022sim2testbed,ni2015named,nurkahfi2021design}, or specific research-related testbed~\cite{lim2018ndn,LTE2019testbed,4G2015testbed}. We categorized all these works into three areas: i) infrastructure emphasized testbeds to implement different scenarios in wired~\cite{sdn2023adaptive,susilo2023forwarding}, ISM band wireless~\cite{saxena2017implementation}, Cellular 4G infrastructure~\cite{4G2015testbed,LTE2019testbed}, edge network~\cite{li2022icast}, a Virtual Machine(VM) environment~\cite{bardhi2022sim2testbed}. All of the related works can run scenarios in a single infrastructure, in addition, adding, deleting, and configuring new nodes is manual and makes challenges for running nodes scenario especially when there are several nodes in the scenario. ii) application specific, most of research can implement the specific research topic software that should be installed on every nodes\cite{lim2018ndn}, so running different applications on testbed needs many time to configure nodes, iii) and performance monitoring testbeds is one of the important types of testbeds which shows different metric outputs of an experimental setup~\cite{lailari2015experiments,lim2018ndn,rainer2016low}. The biggest challenge on this part of a testbed is how to gather several types of logs from different log sources in the central module and run queries on raw logs.
In terms of infrastructure, most testbeds just focus on specific OS platforms, infrastructure, fixed topology, manual routing, and a fixed number of nodes. 

In this paper we propose \sol as a general purpose testbed that implements different scenarios on Ethernet 802.3 wired network~\cite{8023}, or wireless 802.11x ad-hoc networks~\cite{han2004wireless} regardless of hardware and OS platforms. In addition, to make an easily replicable testbed \sol makes a fully automated node's discovery mechanism to add, configure, and delete from the network topology. Users of \sol can exploit this mechanism to install, and update their application in all network nodes. 

However, \sol an application updating methodology lets users develop and automatically deploy applications in testbed nodes, this process is fast and relies on robust node management tools. 

Finally, \sol exploits a powerful centralized logging module to receive different types of logs and generate performance monitoring plots for users. The evaluation module then creates different types of plots based on received logs.

%Nmae Data Networking \cite{bardhi2022sim2testbed} attacks on ndn implementation of ndn.
%Rainer \cite{rainer2016low}just measurement, and evaluation on 7 node.
%resource utilization linux \cite{nurkahfi2021design} also NLSR
%\cite{lim2018ndn} application climate modeling and caching
%\cite{lailari2015experiments} application evaluation IP ndn.
%\cite{saxena2017implementation} vndn simuklation
%\cite{4G2015testbed,LTE2019testbed} 4G test bed
%\cite{susilo2023forwarding} Mini NDN application
%\cite{sdn2023adaptive}
%\cite{ni2015named} connecting to global network.

\section{Named Data Networking (NDN) Overview}\label{NDN}

NDN is an innovative architecture to Information-Centric Networking (ICN) that focuses on the name objects itself rather than the locations of data (IP addresses) as in traditional networking~\cite{jacobson2014named}. NDN is designed around the concept of naming data, which allows it to be directly requested and retrieved based on its content rather than its location. Unlike traditional IP-based networks, NDN is based on a pull-based concept, an interest packet from the consumer side requests content from producers, and a corresponding data message is sent to the consumer in return. This paradigm shift introduces several key components: the Forwarding Information Base (FIB), Pending Interest Table (PIT), and Content Store (CS).

\begin{itemize}
      \item \textbf{Forwarding Information Base (FIB):} is similar to a routing table in IP-based networks, but instead of mapping IP addresses to interfaces, it maps name prefixes to network interfaces. When an interest packet (a request for data) is received, the FIB directs the request towards the most likely direction where the data can be found, ensuring efficient data retrieval.
    \item \textbf{Pending Interest Table (PIT):} temporarily stores information about interests that have been forwarded but not yet satisfied by data. This table keeps track of which interfaces have sent out requests for specific data, allowing the network to deliver the data back to all interested parties when it becomes available, thus enabling multicast-like behavior naturally. Moreover, all interests of the same name prefix will be aggregated on the first PIT record.
    \item \textbf{Content Store (CS):} acts as a cache at the router level, storing copies of recently retrieved data. When a router receives an interest packet, it first checks the CS to see if it has the requested data cached. If it does, the data is sent back immediately, reducing latency and bandwidth usage by avoiding the need to fetch the data from the source again
\end{itemize}

 Forwarding strategies in NDN play a crucial role in determining how interest packets are forwarded through the network to retrieve the desired data. Unlike traditional IP networks where routing is based on static paths defined by IP addresses, NDN's forwarding strategies are more dynamic and content-aware, taking advantage of multiple paths and real-time network conditions. NDN employs various forwarding strategies to efficiently route interest packets to the correct data source. Best route, multicast \cite{tariq2019forwarding}, and Self Learning\cite{shi2017broadcast} are some important forwarding strategies in NDN.

\section{Testbed Overview}\label{testbed overview}

\begin{figure}[tbp]
    \centering
    \includegraphics[width=0.8\linewidth]{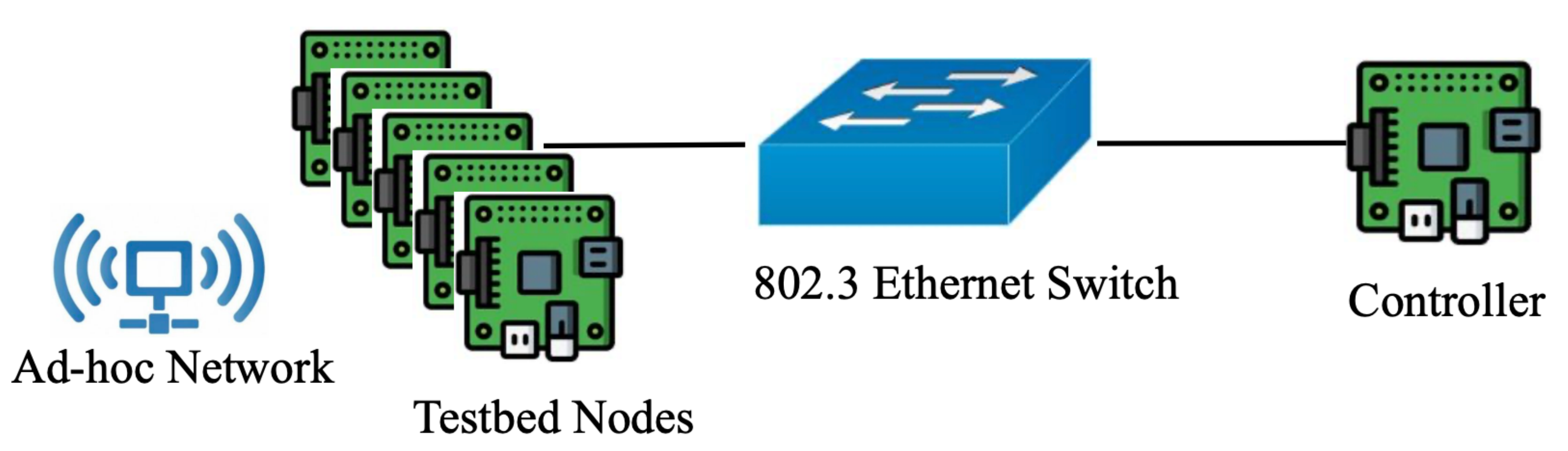}
    \caption{\sol physical topology includes a controller, and nodes, as well as two wired, and wireless network infrastructures.}
    \label{physical}
\end{figure}

In this section, we explain the whole overview of our testbed. \sol includes a controller, an Ethernet switch, and several nodes (Figure~\ref{physical}). The controller is a software package that runs on an independent, or one of the testbed nodes. A switch is needed for wired connections, and nodes can be desktop computers, Raspberry Pi, or even VMs. 

The network infrastructure of \sol includes two different mediums: i) wired Ethernet based on IEEE 802.3~\cite{8023} protocol in this segment all nodes are connecting to the same local area network (LAN) via an Ethernet switch, ii) wireless networks that all nodes are connecting to the ad-hoc wireless network~\cite{han2004wireless}. It is worth mentioning \sol's wireless infrastructure can be also based on a centralized Access Point (AP) network. Each node should have two network interface cards to connect to two infrastructures. Generally, the wired network is used for network management communications, and \sol does not depend on an out-of-band management network. Furthermore,  both infrastructures can be activated at the same time if source routing becomes enabled~\cite{boushaba2014source}.

\subsection{Controller Architecture}

\begin{figure}[tbp]
    \centering
    \includegraphics[width=1.0\linewidth]{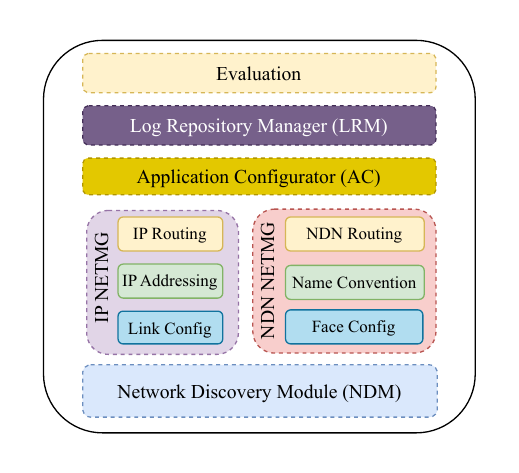}
    \caption{\sol controller architecture.}
    \label{Arcitechture}
\end{figure}

\sol controller package designed based on a modular architecture~\cite{NETSCANDN}. it consists of five independent modules (Figure \ref{Arcitechture}). All these modules were developed by Python3~\cite{zadka2019paramiko}, and C++ programming languages. Four of these modules are common for NDN, and IP protocol stack, and one of them is different for NDN, and IP. In the next sections, we explain about each module.

\subsubsection{Network Discovery Module (NDM)}\label{NDM}

The paramilitary module of \sol is the Network Discovery Module (NMD). While \sol is an easily replicable testbed, this module follows an automatic process to discover all available nodes and configure them for running experiments. This module includes two phases:

\textbf{Initializing Phase:} Our testbed controller runs a Dynamic Host Configuration Protocol (DHCP)~\cite{rfc2131} service on the wired infrastructure, once an IPv4 address is assigned, a node sends its OS type using Vendor Class Identifier (VCI-Option 60)~\cite{rfc2132}, and NDM records OS type on DHCP lease file. Later, it will use this information for the pre-configuration phase.  

\textbf{Pre-Configuring Phase:} In this phase, NDM exploits Ansible~\cite{hochstein2017ansible} (an automatic remote node management package) to install all software requirements of a node. The list of nodes is available in the DHCP lease file, and the only needed information for each node is Secure Shell (SSH)~\cite{ssh2006secure} credential that should be already set on the node. The process starts with the output of the last phase, and based on each node OS a playbook deployed on the node by controller's Ansible software. We implement some playbooks for Ubuntu, MAC, and Pi OS, however, the playbook can be customized for any other OS.

To run the NDN protocol stack on each node we use Named data networking Forwarding Daemon (NFD)~\cite{jacobson2014named}, finally, after deploying all requirements list of ready nodes is sent to the next module for configuring the network topology and other requirements.

\subsubsection{Network Management Module (NETMG)}

NETMG configures the network infrastructure. It remotely installs network configurations via Ansible playbooks on each node. This module includes two independent sub-modules for IP, and NDN respectively: 

\textbf{IP NETMG:} This sub-module includes three phases, i) Link Config that activates the requested interface based on wired and wireless infrastructure, ii) IP addressing assigns IPv4 address to the interface, iii) IP routing that sets routing configuration to the node. 

\begin{figure}[tbp]
    \centering
    \includegraphics[width=1.0\linewidth]{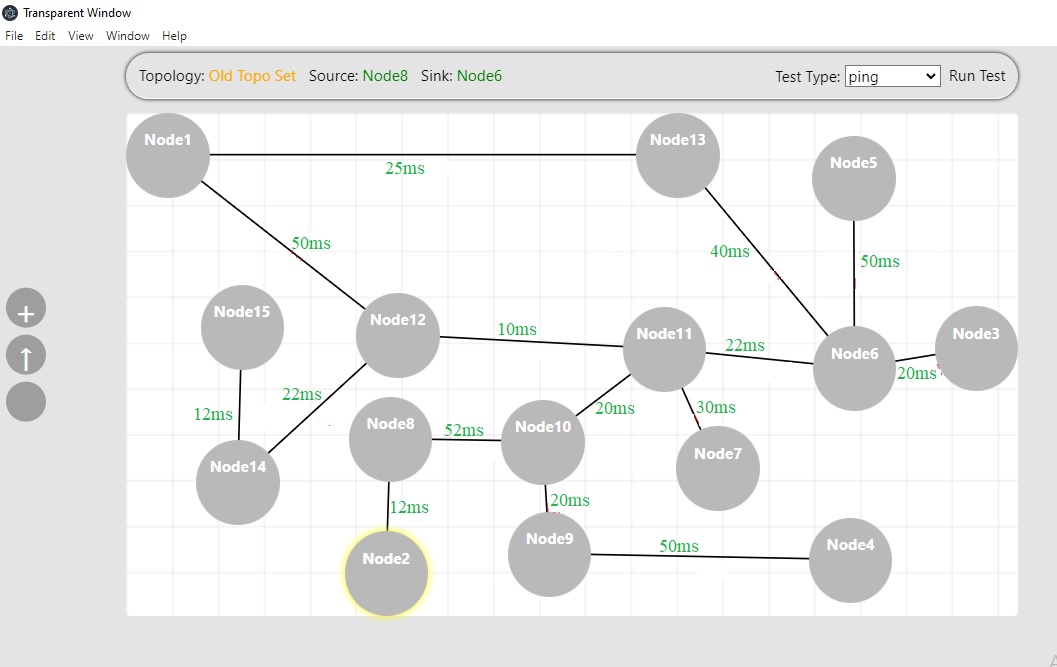}
    \caption{User defined topology via \sol's GUI includes 15 nodes. Link delays are automatically measured and shown in GUI}
    \label{topo}
\end{figure}

IP NETMG receives a network topology from the user side, user sends the requested topology in the format of adjacency matrix~\cite{adjacency} via either testbed Graphical User Interface (GUI) or API (Figure~\ref{topo}). The GUI is designed by ElectronJs~\cite{electronjs}, and the user can manually design the topology. In addition, it can directly receive the topology configuration via RESTFUL API~\cite{richardson2013restful}.

\sol exploits two strategies for IP Routing sub-module for a given network topology: First, global routing when there is no routing protocol, in this case, IP Routing uses Dijkstra~\cite{dijkstra} algorithm to find the best path, and set the corresponding next hop for each destination\footnote{We set a routing prefix for each node, so if there are $N$ node in the testbed, there are $N-1$ destination for each node. Each routing prefix is a /32 prefix.}. Second, if dynamic routing protocol is required for an experiment, \sol uses Bird~\cite{filip2010bird} daemon to run either Open Shortest Path (OSPF)~\cite{ospf}, or Border Gateway Protocol (BGP)~\cite{bgp}, based on user request. This module also can be used for either distributed load balancing methodologies~\cite {esmaeili2018new,hoque2023ecomves}, or creating customized topologies based on network requirements \cite{esmaeili2024samba,esmaeili2024serene}.

\textbf{NDN NETMG:} After configuring the IP routing of nodes, this sub-module configures NFD for NDN routing. 
First, Face Config creates corresponding faces to neighbor nodes. \sol Face Config can create several types of NDN face (Ethernet, TCP, UDP, and web socket). It creates point-to-point links between nodes.
The second step toward configuring the NDN  protocol stack is name convention, we allocate a main prefix for each node $n_i$ so-called $/testbed/P_i$, and in the Application Configuration (AC) module it can be divided into more name prefixes in a hierarchical architecture and according to test scenario. 

Finally, NDN Routing generates and installs NDN routes for each destination. If static routing is called, global routing sets static routes on each node same as IP routing policies, otherwise, if IGP  routing is called, NLSR~\cite{nlsr} will be used.

\subsubsection{Application Configuration (AC)}

\sol as a scalable general multi-purpose testbed can run a variety of applications. Applications run over either IP or NDN protocol stack. There are some pre-installed IP applications for tracing, and measuring IP traffic~\cite{hsu1998iperf}, also for the NDN protocol stack we use NDN Essential tools~\cite{ndnping} including several application testing tools.

However, to run customized software on testbed nodes, we use an automatic deployment approach that is used in section~\ref{NDM}, and using Ansible playbooks that are designed for application deployment, all software prerequisites, and main applications will be installed, and updated on all testbed nodes. 

Additionally, there are two ways to access each node. The first method of accessing each node and running some commands is via SSH remote access, this method is mostly used in cases of troubleshooting, the second that used mostly when there is a scenario to run concurrently on a bunch of nodes, we run a playbook includes needed commands on corresponding nodes. Ansible will show the output  of commands in the terminal, or a log file.

\subsubsection{Log Repository Manager (LRM)}\label{LRM}
While running different applications and approaches create a variety of logs, to measure and evaluate the performance of running scenarios the controller needs to integrate all raw logs in a central repository. \sol's controller exploits syslog~\cite{gerhards2009syslog} which is an IP-based protocol to integrate raw logs from network elements.

Each node sends raw logs to the central syslog server in the controller via UDP port number 514, all these communications are established in a wired network. Moreover, all logs in the NFD application are sent to the controller in the same way. The controller stores all raw logs in a central repository, and logs can be distinguished based on nodes IP address. Also, every log record will be tagged by a timestamp for the performance evaluation and future measurement accuracy. To this end, The controller runs a time server and nodes synchronize the local time with the network time using Network Time Protocol (NTP)~\cite{ntp}.  

However, the syslog protocol has 8 different severity levels, and users can increase/decrease the level of logging based on their requirements, and create a larger, or smaller set of raw logs. Syslog client automatically sends a range of requested severity to the central log module.

\subsubsection{Evaluation}\label{Evaluation}
One of the most important sections of each testbed is performance evaluation. It measures and shows the results of experiments based on defined metrics. The larger the set of metrics, the better the testbed performance evaluation. \sol Evaluation module works based on raw logs received in the central log repository. 
Metrics have a major role in showing the results of experiments. We categorized measurable metrics on \sol into two categories of metrics: 
 
 \paragraph{\textbf{General Metrics}} The First set of metrics indicates the results of running  applications on testbed nodes. \sol evaluation module automatically measures some of the frequently used metrics: i) Throughput is the first metric that is measured based on received interests on each node, ii) interest Round Trip Time (RTT) is the next automatic metric that is measured on evaluation module, iii) link delay is other metric that shows the real-time point to point delay between nodes in its GUI. The evaluation module uses Ping~\cite{ping} and ndnPing~\cite{ndnping} for IP and NDN respectively to measure the RTT of each link between nodes periodically every five seconds (Figure~\ref{topo}). 
    
     \paragraph{\textbf{Node  Related Metrics}} The next type of metrics shows the resource utilization of testbed nodes. We use ifstat~\cite{ifstat} to measure the performance of each NIC, moreover, CPU, RAM, and storage utilization are some other metrics that can be measured from all nodes.  

However, while LRM (section~\ref{LRM}) stores all types of logs, as a flexible testbed \sol's users can create any user-specific logs in their application, send them via syslog protocol to the LRM, and run any required queries to measure user-defined metrics that do not exist in the evaluation section. 
\begin{figure}
    \centering
    
    \subfloat[A simple topology includes 4 routers and a consumer, shows effect of link failure]{\includegraphics[width=0.60\linewidth]{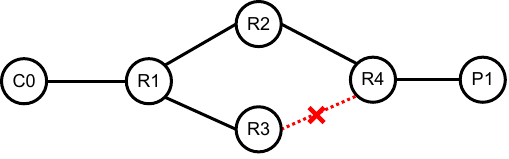}
        \label{fw-topo}}
    \hfill
    \subfloat[The throughput of C0 consumer degrades when Best Route is used, while Multicast strategy is resilient to failure.]{\includegraphics[width=0.60\linewidth]{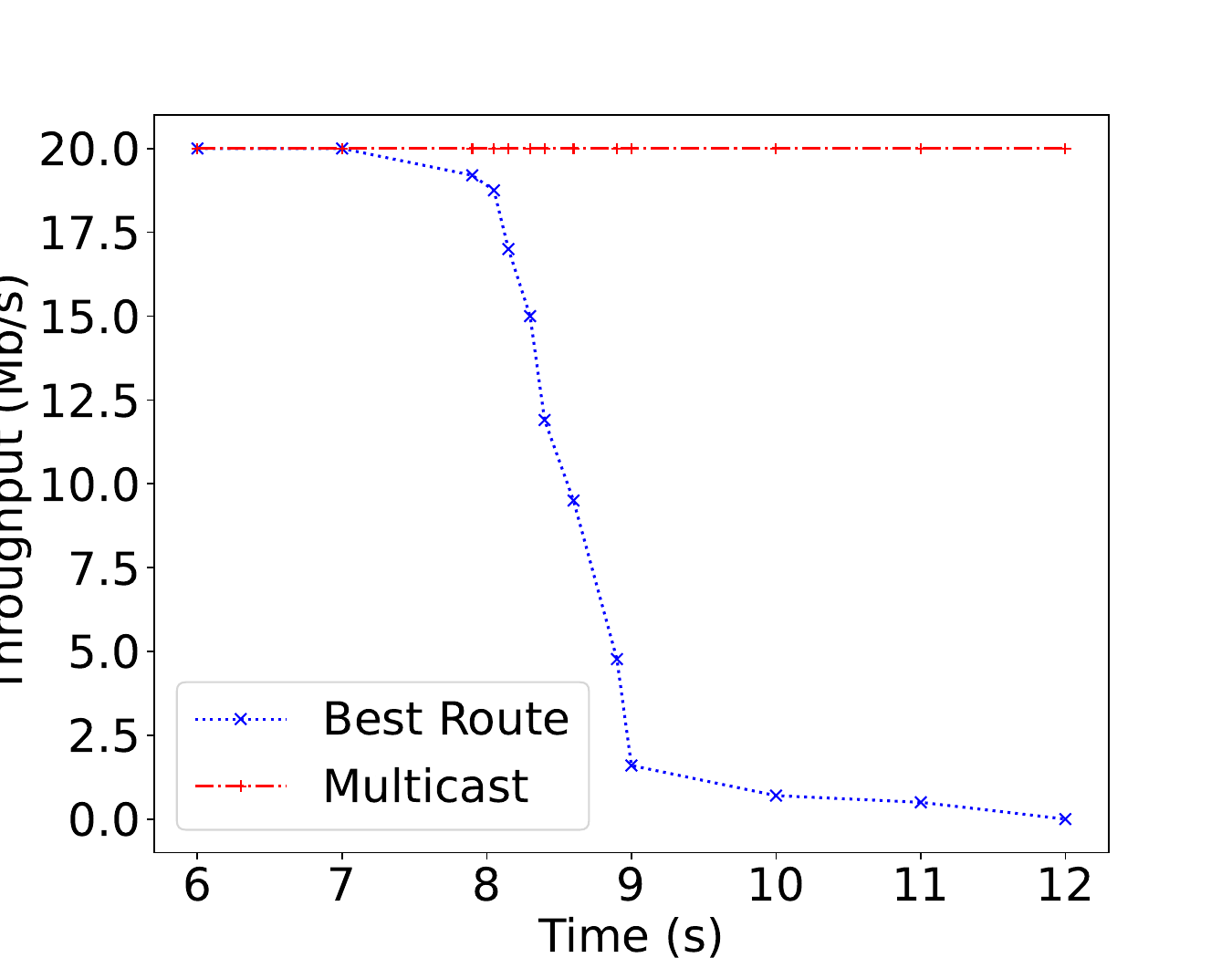}
        \label{fw-throughput}}
    
    \caption{In this experiment, the testbed uses the best path and multicast forwarding strategies and results show link degrading throughput of C0 on link failure when the best route is used, also multicast forwarding strategy is resilient against link failure.}
    \label{visualize}
\end{figure}

\begin{figure*}[tbp]
    \centering  
   \subfloat[Memory and CPU utilization]
     {\includegraphics[width=0.35\linewidth]{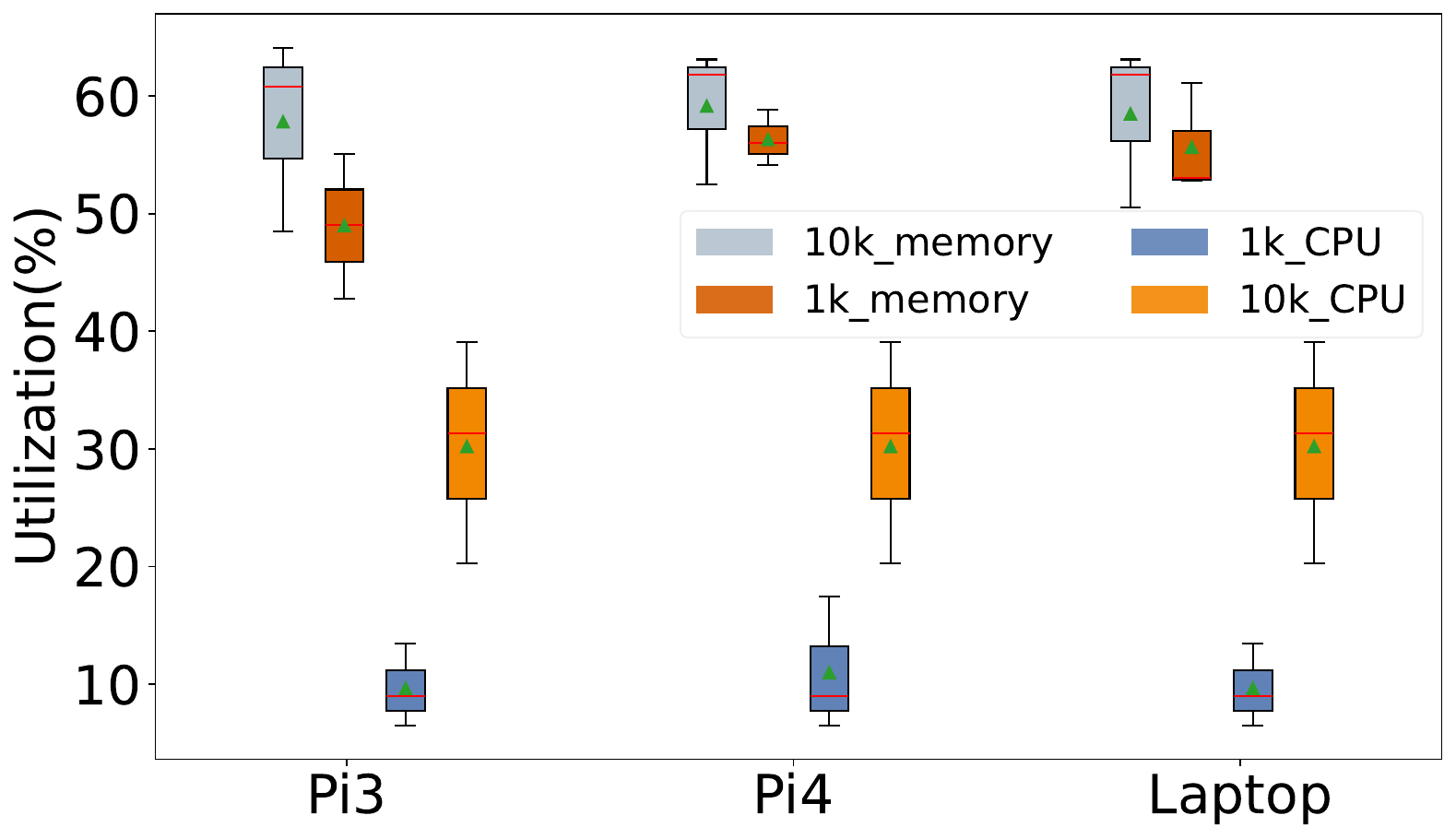}
    \label{benchmarking-Run}}
     \subfloat[Run times]
     {\includegraphics[width=0.35\linewidth]{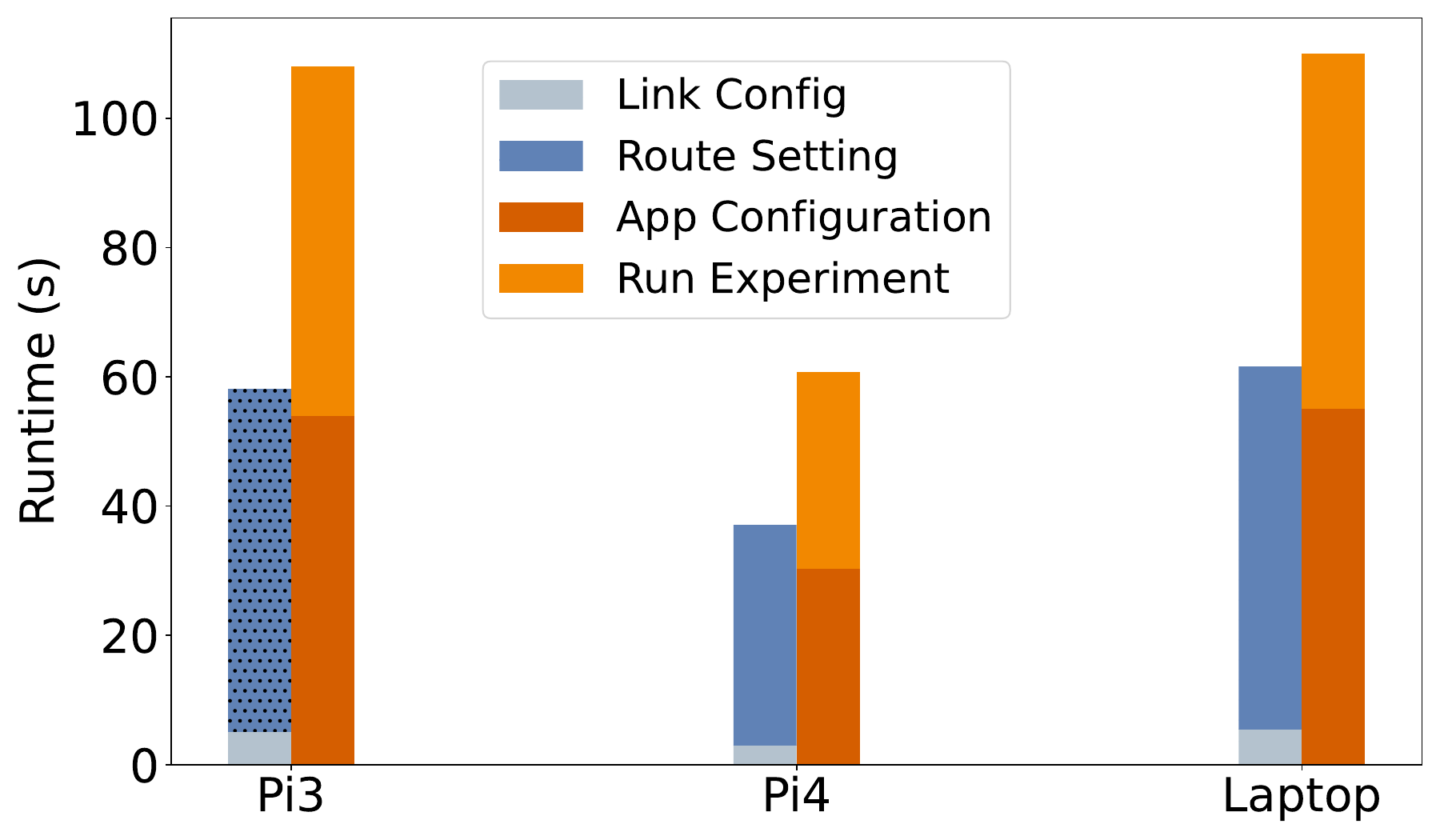} \label{benchmarking_CPU}}
    \caption{The benchmarking tests for memory and CPU utilization (a) as well as run times (b) using a Raspberry Pi3, Raspberry Pi4, and a Laptop machine. 
    }
   \label{benchmarking}
\end{figure*}

\paragraph{Visualization}
Visualization is the last part of the evaluation module, to visualize the result of measured metrics. We implement two types of line plots in the evaluation module using python matplotlib library~\cite{tosi2009matplotlib}.

\noindent{\bf A Simple Configured Scenario:}
In this experiment, We show a link failure scenario with a simple topology shown in Figure~\ref{fw-topo}. The network topology consists of four routers, and in the second 8, the link between R3-R4 fails, we run the experiment 20 times, and the interest sending rate is 20 Mb/sec, the consumer uses AIMD congestion control~\cite{schneider2016practical} to control traffic flow. The C0 consumer requests a content from P1 producer, 
we compare the performance of the best route, and multi-cast forwarding strategy~\cite{tariq2019forwarding} when link failure happens, and as depicted in Figure~\ref{fw-throughput}, the multicast strategy is resilient to link failure. The evaluation module received logs of sending interests and measured the throughput of the consumer when the best route and multicast forwarding strategy were used respectively. 

Moreover, in another experiment, we measure node-related metrics.
This benchmarking experiment includes 10 nodes, the controller configures 10 nodes with 1k and 10k different name prefixes. A testbed user just connects nodes to the wired infrastructure, then runs the controller module, it automatically discovers nodes and also detects the OS type. Nodes required applications installed via Ansible, then a network topology includes 1k, and 10k routing prefix deployed in the node set.  

Figure~\ref{benchmarking-Run} shows the average used CPU and memory in three different node types. These nodes are Raspberry Pi3, Raspberry Pi4, and laptop nodes, and we can see the additional load for configuring 1k, and 10k prefixes in terms of installing routing prefixes. We use global routing which calculate each node next hop automatically, then install configurations on each node.

Finally, we measure the average runtime of different controller modules in a topology that includes 10 nodes (Figure~\ref{benchmarking_CPU}). The Link Config is very fast because of setting just some addresses to each node, and App Config, and Routing (IP, and NDN) are the slowest modules due to heavy duties. The performance of these modules depends on the hardware specification of the OS. Raspberry Pi4 also is the best platform among all other tested platforms.

\section{Conclusion, Future Works}\label{conclusion}

In this paper, we presented \sol a scalable, flexible and plug and play testbed for implementation of IP/NDN applications on different wired, wireless, or both infrastructures simultaneously. \sol exploits an automated process to discover, configure, and run different user-defined applications on several independent nodes. In addition, it can run complex network topologies in both wired and wireless infrastructures, the global routing module helps to configure each node routing configuration in IP and NDN protocols. Finally, it exploits a central log management module to store and measure user-defined metrics for running experiments. 

In the future we will move forward to add more physical mediums to our testbed to make it more flexible for all wireless environments, also adding more forwarding strategies to the list of available forwarding strategies. Finally, we will make an interconnection between our testbed published version, and global NDN testbed.

\bibliographystyle{plain}
\bibliography{main}

\begin{thebibliography}{10}

\bibitem{NETSCANDN}
{A. Esmaeili}.
\newblock {\em NetScaNDN}, 2023.
\newblock Available at: \url{https://github.com/ae3wc/NDNTESTBED}.

\bibitem{rfc2132}
S.~Alexander and R.~Droms.
\newblock Dhcp options and bootp vendor extensions.
\newblock Request for Comments, 1997.

\bibitem{bardhi2022sim2testbed}
Enkeleda Bardhi, Mauro Conti, Riccardo Lazzeretti, Eleonora Losiouk, and Ahmed Taffal.
\newblock Sim2testbed transfer: Ndn performance evaluation.
\newblock In {\em Proceedings of the 17th International Conference on Availability, Reliability and Security}, pages 1--9, 2022.

\bibitem{LTE2019testbed}
Pedro Batista, Ivanes Ara{\'u}jo, Neiva Linder, Kim Laraqui, and Aldebaro Klautau.
\newblock Testbed for icn media distribution over lte radio access networks.
\newblock {\em Computer Networks}, 150:70--80, 2019.

\bibitem{boushaba2014source}
Mustapha Boushaba, Abdelhakim Hafid, and Michel Gendreau.
\newblock Source-based routing in wireless mesh networks.
\newblock {\em IEEE systems Journal}, 10(1):262--270, 2014.

\bibitem{bgp}
Matthew Caesar and Jennifer Rexford.
\newblock Bgp routing policies in isp networks.
\newblock {\em IEEE network}, 19(6):5--11, 2005.

\bibitem{8023}
Ken Christensen, Pedro Reviriego, Bruce Nordman, Michael Bennett, Mehrgan Mostowfi, and Juan~Antonio Maestro.
\newblock Ieee 802.3 az: the road to energy efficient ethernet.
\newblock {\em IEEE Communications Magazine}, 48(11):50--56, 2010.

\bibitem{sdn2023adaptive}
Vassilis Demiroglou, Lefteris Mamatas, and Vassilis Tsaoussidis.
\newblock Adaptive ndn, dtn and nod deployment in smart-city networks using sdn.
\newblock In {\em 2023 IEEE 20th Consumer Communications \& Networking Conference (CCNC)}, pages 1092--1097. IEEE, 2023.

\bibitem{ping}
Danny Dolev, Shimon Even, and Richard~M Karp.
\newblock On the security of ping-pong protocols.
\newblock {\em Information and Control}, 55(1-3):57--68, 1982.

\bibitem{rfc2131}
R.~Droms.
\newblock Dynamic host configuration protocol.
\newblock Request for Comments, 1997.

\bibitem{electronjs}
{Electron Project}.
\newblock {\em Electron: Build cross-platform desktop apps with JavaScript, HTML, and CSS}.
\newblock OpenJS Foundation, 2023.
\newblock Available at: \url{https://www.electronjs.org/}.

\bibitem{ospf}
M~Ericsson, Mauricio G.~C. Resende, and Panos~M. Pardalos.
\newblock A genetic algorithm for the weight setting problem in ospf routing.
\newblock {\em Journal of combinatorial optimization}, 6:299--333, 2002.

\bibitem{esmaeili2018new}
A~Esmaeili and B~Bakhshi.
\newblock A new bgp-based load distribution approach in geographically distributed data centers.
\newblock {\em Nashriyyah-i Muhandisi-i Barq va Muhandisi-i Kampyutar-i Iran}, 62(1):71--78, 2018.

\bibitem{esmaeili2024samba}
Amir Esmaeili and Abderrahmen Mtibaa.
\newblock Samba: Scalable approximate forwarding for ndn implicit fib aggregation.
\newblock {\em arXiv preprint arXiv:2409.19154}, 2024.

\bibitem{esmaeili2024serene}
Amir Esmaeili and Abderrahmen Mtibaa.
\newblock Serene: A collusion resilient replication-based verification framework.
\newblock {\em arXiv preprint arXiv:2404.11410}, 2024.

\bibitem{filip2010bird}
Ondrej Filip, L~Forst, P~Machek, M~Mares, and O~Zajicek.
\newblock Bird internet routing daemon.
\newblock {\em NANOG-48, Austin, TX}, 2010.

\bibitem{gerhards2009syslog}
Rainer Gerhards.
\newblock The syslog protocol.
\newblock Technical report, 2009.

\bibitem{4G2015testbed}
Han.
\newblock {\em A Testbed for Mobile Named-Data Network integrated with 4G networking devices}.
\newblock PhD thesis, 2015.

\bibitem{han2004wireless}
Lu~Han.
\newblock Wireless ad-hoc networks.
\newblock {\em Wireless Personal Communication Journal of mobile communication and computing}, 4, 2004.

\bibitem{hochstein2017ansible}
Lorin Hochstein and Rene Moser.
\newblock {\em Ansible: Up and Running: Automating configuration management and deployment the easy way}.
\newblock " O'Reilly Media, Inc.", 2017.

\bibitem{nlsr}
AKM~Mahmudul Hoque, Syed~Obaid Amin, Adam Alyyan, Beichuan Zhang, Lixia Zhang, and Lan Wang.
\newblock Nlsr: Named-data link state routing protocol.
\newblock In {\em Proceedings of the 3rd ACM SIGCOMM workshop on Information-centric networking}, pages 15--20, 2013.

\bibitem{hoque2023ecomves}
Sanzida Hoque.
\newblock ecomves: Enhancing comves using data piggybacking for resource discovery at the network edge.
\newblock 2023.

\bibitem{hsu1998iperf}
Chung-Hsing Hsu and Ulrich Kremer.
\newblock Iperf: A framework for automatic construction of performance prediction models.
\newblock In {\em Workshop on Profile and Feedback-Directed Compilation (PFDC), Paris, France}. Citeseer, 1998.

\bibitem{ndnping}
{J. Abraham}.
\newblock {\em NDN Essential Tools}, 2019.
\newblock Available at: \url{https://github.com/named-data/ndn-tools}.

\bibitem{jacobson2014named}
Van Jacobson, Jeffrey Burke, Lixia Zhang, Beichuan Zhang, Kim Claffy, Dmitri Krioukov, Christos Papadopoulos, Tarek Abdelzaher, Lan Wang, Edmund Yeh, et~al.
\newblock Named data networking (ndn) project 2013-2014 report.
\newblock {\em Palo Alto Research Centre}, 2014.

\bibitem{lailari2015experiments}
Ze'ev Lailari, Hila~Ben Abraham, Ben Aronberg, Jackie Hudepohl, Haowei Yuan, John DeHart, Jyoti Parwatikar, and Patrick Crowley.
\newblock Experiments with the emulated ndn testbed in onl.
\newblock In {\em Proceedings of the 2nd ACM Conference on Information-Centric Networking}, pages 219--220, 2015.

\bibitem{li2022icast}
Tianlong Li, Tian Song, Yating Yang, and Jike Yang.
\newblock icast: dynamic information-centric cross-layer multicast for wireless edge network.
\newblock In {\em Proceedings of the 9th ACM Conference on Information-Centric Networking}, pages 137--147, 2022.

\bibitem{lim2018ndn}
Huhnkuk Lim, Alexander Ni, Dabin Kim, Young-Bae Ko, Susmit Shannigrahi, and Christos Papadopoulos.
\newblock Ndn construction for big science: Lessons learned from establishing a testbed.
\newblock {\em IEEE Network}, 32(6):124--136, 2018.

\bibitem{ifstat}
{M. Baerts}.
\newblock {\em ifstat Tools}, 2020.
\newblock Available at: \url{https://github.com/matttbe/ifstat}.

\bibitem{ntp}
David Mills, J~Burbank, and W~Kasch.
\newblock Rfc 5905: Network time protocol version 4: Protocol and algorithms specification, 2010.

\bibitem{ni2015named}
Alexander Ni and Huhnkuk Lim.
\newblock A named data networking testbed with global ndn connection.
\newblock {\em tre}, 40(12):2419--2426, 2015.

\bibitem{dijkstra}
Masato Noto and Hiroaki Sato.
\newblock A method for the shortest path search by extended dijkstra algorithm.
\newblock In {\em Smc 2000 conference proceedings. 2000 ieee international conference on systems, man and cybernetics.'cybernetics evolving to systems, humans, organizations, and their complex interactions'(cat. no. 0}, volume~3, pages 2316--2320. IEEE, 2000.

\bibitem{nurkahfi2021design}
Galih~Nugraha Nurkahfi, Tody~Ariefianto Wibowo, Syaiful Ahdan, Arumjeni Mitayani, Mochamad Mardi~Marta Dinata, Vita~Awalia Mardiana, Nana~Rachmana Syambas, Ade Nurhayati, and Hafizh~Mulya Harjono.
\newblock Design and implementation of low-cost and highly customable named data networking (ndn) testbed.
\newblock In {\em 2021 15th International Conference on Telecommunication Systems, Services, and Applications (TSSA)}, pages 1--7. IEEE, 2021.

\bibitem{rainer2016low}
Benjamin Rainer, Daniel Posch, Andreas Leibetseder, Sebastian Theuermann, and Hermann Hellwagner.
\newblock A low-cost ndn testbed on banana pi routers.
\newblock {\em IEEE Communications Magazine}, 54(9):105--111, 2016.

\bibitem{richardson2013restful}
Leonard Richardson, Mike Amundsen, and Sam Ruby.
\newblock {\em RESTful Web APIs: Services for a Changing World}.
\newblock " O'Reilly Media, Inc.", 2013.

\bibitem{adjacency}
Ronald~L Rivest and Jean Vuillemin.
\newblock On recognizing graph properties from adjacency matrices.
\newblock {\em Theoretical Computer Science}, 3(3):371--384, 1976.

\bibitem{saxena2017implementation}
Divya Saxena, Vaskar Raychoudhury, and Christian Becker.
\newblock Implementation and performance evaluation of name-based forwarding schemes in v-ndn.
\newblock In {\em Proceedings of the 18th International Conference on Distributed Computing and Networking}, pages 1--4, 2017.

\bibitem{schneider2016practical}
Klaus Schneider, Cheng Yi, Beichuan Zhang, and Lixia Zhang.
\newblock A practical congestion control scheme for named data networking.
\newblock In {\em Proceedings of the 3rd ACM Conference on Information-Centric Networking}, pages 21--30, 2016.

\bibitem{shi2017broadcast}
Junxiao Shi, Eric Newberry, and Beichuan Zhang.
\newblock On broadcast-based self-learning in named data networking.
\newblock In {\em 2017 IFIP Networking Conference (IFIP Networking) and Workshops}, pages 1--9. IEEE, 2017.

\bibitem{susilo2023forwarding}
Reza~Maharani Susilo, Farraz~Rizky Kusumaputra, Muhammad~Hendrawan Adiwijaya, Ratna Mayasari, Ridha~Muldina Negara, and Sri Astuti.
\newblock Forwarding strategy analysis in wireless network based named data network (ndn).
\newblock In {\em 2023 International Conference on Electrical and Information Technology (IEIT)}, pages 361--366. IEEE, 2023.

\bibitem{tariq2019forwarding}
Asadullah Tariq, Rana~Asif Rehman, and Byung-Seo Kim.
\newblock Forwarding strategies in ndn-based wireless networks: A survey.
\newblock {\em IEEE Communications Surveys \& Tutorials}, 22(1):68--95, 2019.

\bibitem{tosi2009matplotlib}
Sandro Tosi.
\newblock {\em Matplotlib for Python developers}.
\newblock Packt Publishing Ltd, 2009.

\bibitem{ssh2006secure}
Tatu Ylonen and Chris Lonvick.
\newblock The secure shell (ssh) connection protocol.
\newblock Technical report, 2006.

\bibitem{zadka2019paramiko}
Moshe Zadka and Moshe Zadka.
\newblock Paramiko.
\newblock {\em DevOps in Python: Infrastructure as Python}, pages 111--119, 2019.

\bibitem{zhang2014named}
Lixia Zhang, Alexander Afanasyev, Jeffrey Burke, Van Jacobson, KC~Claffy, Patrick Crowley, Christos Papadopoulos, Lan Wang, and Beichuan Zhang.
\newblock Named data networking.
\newblock {\em ACM SIGCOMM Computer Communication Review}, 44(3):66--73, 2014.

\end{thebibliography}
\end{document}